\documentclass[fleqn,usenatbib,useAMS]{mnras}

\usepackage{graphicx}	% Including figure files
\usepackage{mwe}
\usepackage{amsmath}	% Advanced maths commands
\usepackage{amssymb}	% Extra maths symbols
\usepackage{multicol}        % Multi-column entries in tables
\usepackage{bm}		% Bold maths symbols, including upright Greek
\usepackage{pdflscape}	% Landscape pages
\usepackage{multirow}
\usepackage{ulem}
\usepackage{comment}
\usepackage{array}
\usepackage{caption}
\usepackage{subcaption}
\usepackage{lineno}
\usepackage{float}
\usepackage{newtxtext,newtxmath}

\includecomment{comment} %show comments
\newcommand{\abs}[1]{\lvert#1\rvert}

\newcommand{\rg}{$r_{\rm g}$}
\usepackage[T1]{fontenc}
\usepackage{ae,aecompl}
\usepackage{newtxtext,newtxmath}
\usepackage{caption}
\usepackage{subcaption}
\usepackage{float}

\title[Lense-Thirring Precession in Truncated Discs]{Effect of geometrically thin discs on precessing, thick flows: Relevance to type-C QPOs}

\author[Bollimpalli et al]{D. A. Bollimpalli$^{1,2,3}$\thanks{Contact e-mail: \href{mailto:deepika@mpa-garching.mpg.de}{deepika@mpa-garching.mpg.de}},
P. C. Fragile$^{3,4}$, W. Klu\'zniak$^{2}$
\\
$^{1}$Max-Planck-Institut f{\"u}r Astrophysik, 85741 Garching b. M{\"u}nchen, Germany\\
$^{2}$Nicolaus Copernicus Astronomical Center, ul. Bartycka 18, PL 00-716 Warsaw, Poland\\
$^{3}$Department of Physics and Astronomy, College of Charleston, Charleston, SC 29424, USA\\
$^{4}$Kavli Institute for Theoretical Physics, Kohn Hall, University of California, Santa Barbara, CA 93107, USA}
\date{}

\pubyear{2020}
\begin{document}
\label{firstpage}
\pagerange{\pageref{firstpage}--\pageref{lastpage}}
\maketitle

\begin{abstract}
Type-C quasi-periodic oscillations (QPOs) are the low-frequency QPOs most commonly observed during the hard spectral state of X-ray binary systems. The leading model for these QPOs is the Lense-Thirring precession of a hot, geometrically thick accretion flow that is misaligned with respect to the black hole spin axis. However, none of the work done to date has accounted for the effects of a surrounding, geometrically thin disc on this precession, as would be the case in the truncated disc picture of the hard state. To address this, we perform a set of GRMHD simulations of truncated discs misaligned with the spin axes of their central black holes. Our results confirm that the inner-hot flow still undergoes precession, though at a rate that is only 5 percent of what is predicted for an isolated, precessing torus. We find that the exchange of angular momentum between the outer, thin and the inner, thick disc causes this slow-down in the precession rate and discuss its relevance to type-C QPOs.
\end{abstract}

%%%%%%%%%%%%%%%%%%%%%%%%%%%%%%%%%%%%%%%%%%%%%%%%%%
%%%%%%%%%%%%%%%%% BODY OF PAPER %%%%%%%%%%%%%%%%%%
\section{Introduction}
\label{sec:intro}
Many accreting black holes and neutron stars exhibit rapid variability in their X-ray light curves, which is often quasi-periodic in nature (indicated by broad peaks of power in the frequency domain). Further, they occasionally undergo outbursts during which the source transits through several different spectral states, as it simultaneously varies in luminosity and spectral hardness, tracing a ``q'' shape in a hardness-intensity diagram, 
%with major shifts being referred to as ``state transitions''
\citep{FBG2004, MR2006}. The current best models that provide a framework for understanding these outbursts relate the changing spectra to a changing geometry of the accretion flow and the presence or absence of different physical components (e.g., a corona or jet) \citep{EMN1997,BA2014}.
The X-ray spectra observed in these systems is mainly composed of a soft, thermal component (below approximately 3 keV) and a hard power-law component (extending roughly between 10 - 100 keV). The soft component is thought to be due to thermal, blackbody-like radiation from a relatively cold, optically thick and geometrically thin disc. The non-thermal, power-law component is thought to be generated from the inverse-Compton scattering of soft seed photons emitted from the disc by a radiatively inefficient cloud of hot electrons, often referred to as the ``corona.'' The location and geometry of the corona are still under much debate, with some models placing it at the base of the jet \citep{MM1996, Fabian2009, Parker2015}, while others consider it to sandwich the disc \citep{GRV1979, HFM1993, S1996, B1999}. The truncated-disc model envisions a third possible geometry with a standard-thin disc truncated well outside the last stable orbit, and the corona as a geometrically thick, radiatively inefficient accretion flow filling in the inner gap \citep{EL1975, EMN1997, LTMM2007}. Although all the above mentioned models can reproduce the observed X-ray spectra, the truncated disc model has the advantage of being able to simultaneously explain some of the observed variability properties. Irrespective of the model, a better understanding of the physical origins of the variability observed in each spectral state can serve as a powerful probe of the inner regions of accretion discs.

The nature of quasi-periodic oscillations (QPOs) varies from one spectral state to another. Broadly speaking, the QPOs observed in black hole systems are classified as high-frequency ($\gtrsim 60$ Hz) and low-frequency ($\lesssim 30$ Hz) \citep{Remillard06, B2010}. Low-frequency QPOs (LFQPOs) are further classified into type-A, B and C, of which the first two are observed in the intermediate states, during the transition from hard to soft spectral states \citep{Motta16}. Type-C QPOs are commonly observed in most accretion states~\citep{Motta12, Motta16}, although they are particularly prominent in the hard state. For these QPOs, both the fractional QPO amplitude and the phase-lag between the hard and soft photons measured at the QPO frequency are found to have an inclination dependence \citep{Motta2015, eijnden2017}. Such a dependence strongly suggests a geometric origin for this QPO. 
%Further support for this argument is provided by the phase-resolved spectroscopic study of a single source, H1743-322, for which the centroid energy of the observed iron line emission is found to vary with the QPO phase \citep{Ingram2016}. 
For a detailed review of type-C QPOs, we refer the reader to \citet{IM2019}.

Most of the models invoked to explain type-C QPOs are based either on the geometry or the intrinsic properties of the disc. Possible mechanisms that have been invoked include trapped corrugation modes ($c$-modes) in the inner regions \citep{KF80, W1999,SWM01} or spiral density waves propagating back and forth between the disc inner radius and the Lindblad radius at some frequency \citep{TP99, RVT2002}. Along the same lines, oscillations of a standing shock in the accretion flow could also produce a type-C QPO; however, to produce the required frequencies, such a shock would need to be located quite far from the black hole \citep{CT95}. Another popular model is the relativistic precession model, which associates the type-C QPO with the nodal precession frequency of a geodesic particle orbit in relativity \citep{SV98, SVM99}. 
% While this model has been widely applied, its strong preference for low spin values has been difficult to reconcile with other measurements \citep{Motta2014}. 

A variation of this model, which seems to match observations of the type-C QPO particularly well, is, instead of having a particle or infinitesimal ring precess, have a finite inner region of the accretion flow precess \citep{Ingram09}. This model ties in with the truncated disc picture of the hard accretion state, in that it is the hot, thick flow inside the truncation radius (i.e., the corona) that is precessing \citep{ID2011}. This has the advantages of lowering the precession frequency into the proper range, naturally explaining the rise in the QPO frequency during the early phases of an outburst cycle (when the truncation radius is moving inward), and easily reproducing the inclination dependence seen in QPO observations (since precession is more apparent the closer to the orbital plane one observes). This model also naturally explains the observed correlation between the QPO phase and the energy of the iron emission lines seen in phase-resolved spectroscopic studies \citep{Ingram2016}. %However, the key assumption of this model is that the inner, hot, geometrically thick flow can precess as a rigid body, independent of the surrounding thin disc.  

Previous numerical work involving magnetohydrodynamical simulations of {\it isolated}, tilted, thick discs have demonstrated global Lense-Thirring precession at frequencies matching observations \citep{Fragile07, Liska18, WQB19}. However, none of those studies consider the full truncated-disc geometry and especially the effects of the surrounding geometrically thin disc on the type-C QPO frequencies. The main concern for QPO models based on the Lense-Thirring precession of the hot, thick flow is its coupling to the outer thin disc. Questions that need to be answered are: 1) How does the surrounding thin disc alter the precession, both in terms of its frequency and amplitude; 2) Will the coupling slow down the precession and eventually stop it or perhaps boost the strength of the QPOs? The main objective of this paper is to address some of these questions. %through a numerical study of tilted, truncated discs surrounding spinning black holes. 

For this purpose, we perform general relativistic magnetohydrodynamics (GRMHD) simulations of tilted, truncated discs around rotating black holes. We evolve the simulations long enough to establish their steady-state behavior which amounts to a significant fraction of the disc's precession period. The remainder of this paper is organized as follows: In Section~\ref{sec:setup}, we describe the initial setup of our simulations. In Section~\ref{results}, we discuss the main result of our paper, that the outer, thin disc slows down the precession of the inner, thick disc. We conclude with a discussion in Section~\ref{discussion}.

\section{Initial Setup }
\label{sec:setup}
\subsection{Physical setup}
We performed three, fully 3D, GRMHD simulations of truncated accretion discs, all with the angular momentum axes of the disc and the black hole misaligned by an angle of $\beta_0 = 15^{\circ}$, as listed in Table~\ref{tab:sims}; one is a high-resolution, 4-level simulation (with dimensionless spin parameter $a_*= a/M = 0.9$), and two are 3-level, low-resolution simulations with $a_*=0.9$ and 0.5. %The initial disc angular momentum vector lies along the $z$-axis, whereas the black hole spin axis is tilted in the $-x$-direction. 

\begin{center}
\begin{table}
\begin{tabular*}{0.5 \textwidth}{ccccc}
\cline{1-5}
Simulation & Spin  &  Tilt & Resolution & $t_\mathrm{end}$ \\
  & ($a_*$) & ($\beta_0$) &  & [\rg$/c$] \\
\cline{1-5}
a9b15r15L4 & 0.9 & $15^{\circ}$ & 384x256x256 & 25,000 \\
%a9b15L4 & 0.9 & $15^{\circ}$ & 384x256x256 & 25,000 & No\\
a9b15r15L3 & 0.9 & $15^{\circ}$ & 192x128x128 & 14,850 \\
a5b15r15L3 & 0.5 & $15^{\circ}$ & 192x128x128  & 16,700 \\
\cline{1-5}
\end{tabular*}
\caption{Simulations presented in this paper.}
\label{tab:sims}
\end{table}
\end{center}

The initial setup of all the simulations consists of a finite torus, surrounded by a thin disc. The torus is initialized following the procedure of \citet{C1985} assuming a constant specific angular momentum, with its inner edge at $r_{\rm in} = 6.5\,$\rg, pressure maximum at $r_{\rm cen} =9\,$\rg, and polytropic index of $\gamma=5/3$, where $r_{\rm g} = GM/c^2$ is the gravitational radius. In all the simulations, a thin slab with fixed height, $H = 0.4$\rg, surrounds the torus extending from $15$\rg to the outer boundary of the simulations domain at $\approx 250$\rg, with a density distribution 
\begin{equation}
\rho(R,z) \propto \frac{\exp(-z^2/2H^2)}{\{1+\exp[(15r_{\rm g}-R)/H]\}\{1+\exp[(R-(40 r_{\rm g})^{1.5})/H]\}} ~,
\end{equation}
where $R=r\sin \theta$ is the cylindrical radius. The angular momentum in the thin disc is Keplerian. In order to maintain the desired thin structure, we use an artificial cooling function, as described in \citet{FW12}, but only applied to radii beyond $15$\rg.

Both the torus and the slab are threaded with numerous small poloidal loops of magnetic field with alternating polarity, starting from a magnetic vector potential of the form
\begin{equation}
    A_{\phi} \propto \frac{1}{1+\exp(\delta)}\sqrt{P_{\rm g}}\sin(2\pi R/5 H) ~,
\end{equation} 
where $P_{\rm g}$ is the local gas pressure and $\delta = 10 [(z/H)^2+ H^2/(R-r_{\rm ms})^2 + H^2/(40r_{\rm g}^{1.5}-R)^2-1]$, with $H = 0.6$ and $0.4$\rg for the torus and thin disc regions, respectively, and $r_{\rm ms}$ is the location of the marginally stable orbit. The resulting magnetic field is normalized such that the ratio of the gas pressure to the magnetic pressure is initially $\ge 10$ throughout the torus and thin disc.% as shown in Fig.~\ref{fig:setup}. 
This field configuration prevents the accumulation of strong net flux in the inner regions, so we do not expect a magnetically arrested disc (MAD) to form. The background region surrounding the thin disc and torus is initialized with a non-magnetic gas with low-density, $\rho = 10^{-5}\rho_{\rm max}r^{-1.5}$ and internal energy density, $e = 10^{-7}e_{\rm max}r^{-2.5}$, where $\rho_{\rm max}$ and $e_{\rm max}$ are the maximum gas density and internal energy density in the disc/torus.
% \begin{figure}
% \begin{subfigure}[t]{0.49\textwidth}
%          \centering
%          \includegraphics[width=\textwidth]{Figures/initial_rho.png}
% \end{subfigure}
% \begin{subfigure}[t]{0.49\textwidth}
%          \centering
%          \includegraphics[width=\textwidth]{Figures/initial_beta.png}
% \end{subfigure}
% \caption{Initial setup. Top: Pseudocolor plot of gas density overlaid with our 4-level mesh. Bottom: Pseudocolor plot of the ratio of magnetic pressure to gas pressure overlaid with magnetic field contours. }
% \label{fig:setup}
% \end{figure}

\subsection{Numerical setup}
All simulations are performed using \textit{Cosmos++} \citep{Anninos05}, with its five-stage, strong-stability-preserving Runge-Kutta (SSPRK) time-stepping scheme \citep{Spiteri02} and third-order, piecewise-parabolic spatial reconstruction \citep{Colella1984}. The fluxes are calculated at zone interfaces using a two-wave, HLL, approximate Riemann solver \citep{Harten1983}.

The simulations are performed in modified Kerr-Schild spherical-polar coordinates, which allow us to place the inner domain boundary inside the black hole event horizon. The grid is logarithmically spaced in $r$, extending over $r\in [1.4 \,r_{\rm g},\,40^{1.5}\,r_{\rm g}]$, and uniformly spaced in $\phi$, covering the full $[0,2\pi]$ domain. The polar angle $\theta$ is uniformly spaced in the coordinate $x_2$, ranging from 0 to 1, with
\begin{equation}
  \theta = \pi x_2 + \frac{1-h}{2} \sin(2 \pi x_2) ~,
\end{equation}
where $h=0.5$ is used to concentrate the grid cells close to the midplane \citep{McKinney06}. We cut out a small cone of opening angle $10^{-15}\pi$ near the poles to avoid calculating metric terms on the pole.

All simulations use a base resolution of $48\times32\times32$ with an additional two or three levels of static mesh refinement added, with adjacent refinement levels differing in resolution by a factor of $2$ in each dimension, to bring most of the grid up to a fiducial resolution of $192\times128\times128$ or $384\times256\times256$ for ``3-level" and ``4-level" simulations, respectively. The initial two levels of refinement are used to resolve both the torus and the thin slab, which is common for both the 3-level and 4-level simulations. The torus region is refined with two refinement levels in the regions within $1.4<r/r_{\rm g}<24$ and $0.14\pi<\theta<0.86\pi$, and $1.4<r/r_{\rm g}<21$ and $0.2\pi<\theta<0.8\pi$, respectively. The thin disc region within $12.6<r/r_{\rm g}<253$ and $0.4\pi<\theta<0.6\pi$ is refined twice. For the 4-level simulation, an additional refinement layer is used to better resolve the thin disc over $r\in[10.5\,r_{\rm g}, 60\,r_{\rm g}]$.

Outflow boundary conditions %\footnote{All fields are copied to ghost zones while ensuring that the velocity component normal to the boundary points outward.}
are used at both the inner and the outer radial boundaries. We use pole-axis boundary conditions near the poles, where information from the corresponding zone across the pole is used for calculating gradients, although fluxes, magnetic fields, and electric fields (emf’s) are zeroed out on the faces and edges that touch the pole. Lastly, periodic boundary conditions are implemented along the azimuthal direction. The zero-divergence of the magnetic fields is ensured using a constrained transport scheme \citep{Fragile12}.
% Whenever density or pressure are boosted as a result of floors and ceilings, the boost is done in the drift-frame in order to preserve momentum along the magnetic field lines \citep{Ressler2017}.

As expected, once the simulations are set to evolve, the initial magnetic fields trigger the magnetorotational instability, causing turbulence to transport the angular momentum and thus allow accretion. The heat generated in this process inflates the torus, thus forming a hot thick flow in the inner regions. The cooling function, which is set to work in the outer thin slab region ($r \ge 15$\rg), acts as a heat sink and helps to maintain the target thickness, $H/r \approx 0.05$. The simulations have been run for more than $25,000$\rg$/c$ for the 4-level simulation and around  $15,000$\rg$/c$ for the 3-level simulations. All of them have successfully maintained a two-component flow, as anticipated.

\section{Results}
\label{results}
\subsection{Precession angle}
\label{sec:precession}
We compute the precession angle, $\gamma$, by adopting the same formalism as in \citet{Fragile07}; this angle quantifies how much the disc angular momentum vector has twisted, or precessed, around the black hole spin axis, as
\begin{eqnarray}
\gamma(r) = \cos^{-1}\left[ \frac{\bm{J}_{\rm BH} \times \bm{J}_{\rm disc}(r)}{\abs{\bm{J}_{\rm BH} \times \bm{J}_{\rm disc}(r)}} \cdot\hat{\textbf{y}}\right] ~,
\label{eq:twist}
\end{eqnarray}
where $\bm{J}_{\rm BH} = \left(-aM \sin \beta_0 \hat{x},\,0,\,aM \cos \beta_0 \hat{z}\right)$ is the angular momentum vector of the black hole, and 
\begin{eqnarray}
\left( \bm{J}_{\rm disc} \right)_{\eta} (r) =  \frac{\epsilon_{\mu \nu \sigma \eta}L^{\mu \nu}S^{\sigma}}{2\sqrt{-S^{\alpha}S_{\alpha}}}
\end{eqnarray}
for $\eta = 1,2,3$ corresponds to the Cartesian vector components of the angular momentum vector of the disc as measured in asymptotically flat space. Here 
\begin{eqnarray}
      L^\mathrm{\mu \nu} = \int\left(x^\mu T^\mathrm{\nu 0} - x^\nu T^\mathrm{\mu 0} \right){\rm d}^\mathrm{3}x  
\end{eqnarray}
and $S^{\sigma} = \int T^{\sigma 0} {\rm d}^\mathrm{3}x$, where $T^{\mu\nu}$ is the MHD stress-energy tensor of the fluid [see Appendix \ref{appA}]. As a reminder, the angular momentum vector of the disc at $t=0$ is aligned along the direction of the $\hat{z}$ unit vector. To avoid the degeneracy in cosine for angles greater than $180^{\circ}$, we also look at the projection of $\bm{J}_{\rm BH} \times \bm{J}_{\rm disc}(r)$ onto $\hat{\textbf{x}}$ . 

\begin{figure}
\begin{subfigure}[t]{0.48\textwidth}
         \centering
         \includegraphics[width=\textwidth]{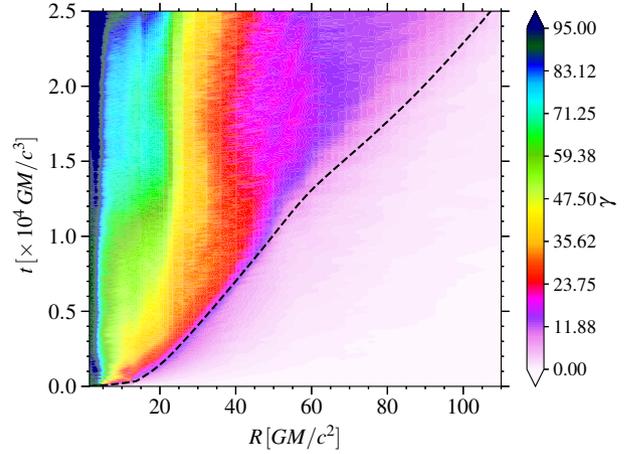}
\end{subfigure}
% \begin{subfigure}[t]{0.48\textwidth}
%          \centering
%          \includegraphics[width=\textwidth]{Figures/precession_rate.png}
% \end{subfigure}
\caption{Space-time plot of the precession angle, $\gamma$ (measured in degrees) %, and the precession rate, $d\gamma/dt$ (bottom panel), 
for simulation a9b15r15L4. The dashed curve represents the bending wave propagating at half the sound speed. The consistent precession angle seen between 5 and 20\rg~at any given time suggest that this region is undergoing rigid-body precession.}
\label{fig:twist}
\end{figure}

Fig.~\ref{fig:twist} shows the evolution of the precession angle for simulation a9b15r15L4. There is a clear signature of a precession front starting from the inner thick-disc region, which propagates to the outer radii (from lower left to upper right) over time. The propagation happens on the bending wave timescale shown by the black, dashed curve \footnote{Bending waves propagate at roughly half the sound speed, $c_{\rm s}$, so the dashed curve is given by $\int 2{\rm d}r/\langle c_{\rm s}(r,t)\rangle_{\theta,\phi}$.}. Radii very close to the black hole undergo rapid precession to angles up to $90^\circ$, holding there for a period until the surrounding regions begin to catch up. Beyond 20 \rg, the initial precession is much more modest ($\gamma \lesssim 30^\circ$). This value also freezes in after a very short period of time ($<2000$ \rg$/c$ after the bending wave passes); this frozen-in twist is noted by the vertical stripes of color in Fig.~\ref{fig:twist}, such as the vertical, red patch at $\approx40$ \rg. Inside of 20 \rg, however, the behavior is much different. Rather than vertical stripes of constant color, we see nearly simultaneous variation of color between 5 and 20 \rg. This simultaneous variation indicates {\it solid-body precession}; all radii are experiencing the same degree of precession at roughly the same time. An example is the nearly horizontal, dark green band around a time of $10^4$ \rg$/c$, indicating that the inner torus has precessed $\approx60^\circ$ by this time.
%The bottom panel of Fig.~\ref{fig:twist} shows the precession rate, $d\gamma/dt$, for the simulation a9b15r15L4. The uniform color in precession rate for the region within $5-15$\rg suggests rigid-body precession of the thick-disc region. Interestingly, when the bending wave reaches the transition radius, the precession rate decreases significantly. This is likely due to the effect of the outer thin disc on the precession of the inner torus, and we discuss this in detail in the following section.

%As the bending wave passes through the thin disc, it is also subjected to precession for a brief period. This happens for two reasons: a) the accretion timescales are longer than the precession timescales in the disc before the bending wave hits, and b) in all our simulations, the viscous-stress parameter \footnote{We compute $\alpha$ as the ratio of the $r$-$\phi$ component of stress in the co-moving frame to the total pressure. Both the stress and the pressure terms include contributions from the magnetic field.}, $\alpha$, remains less than the disc scale height everywhere. The latter could be a consequence of under-resolving the outer thin disc, but the higher resolution simulations needed to test that hypothesis would be quite expensive. However, once the bending wave passes through the thin disc, the accretion timescales fall shorter than the precession timescales. So the thin disc stops to precess eventually and the warps are diffused away by the dynamical processes in the disc.

\subsection{Effect of the outer thin disc on the precessing inner thick disc}
\label{sec:precess}
The Lense-Thirring precession frequency of a test-particle in an orbit misaligned with the spin-axis of the black hole is given by

\begin{equation}
    \Omega_{\rm LT} = \Omega_{\phi}\left[1-\sqrt{1-\frac{4a_*}{r^{3/2}}+\frac{3a_*^2}{r^2}}\right] ~.
    \label{eq:omgLT}
\end{equation}
For an isolated, torus-like structure between an inner and outer radius, $r_{\rm i}$ and $r_{\rm o}$, precessing as a solid body, the angular frequency of precession is computed from weighting $\Omega_{\rm LT}$ by the angular momentum of the torus:
\begin{equation}
    \Omega_{\rm prec} = \frac{\int_{r_{\rm i}}^{r_{\rm o}} \Omega_{\rm LT} \Sigma r^3 \Omega_{\phi} {\rm d} r}{\int_{r_{\rm i}}^{r_{\rm o}} \Sigma r^3 \Omega_{\phi} {\rm d} r} ~,
    \label{eq:prec}
\end{equation}
where $\Omega_{\phi}(r)$ is the local angular velocity. Assuming a power-law profile for the surface density, $\Sigma(r) \sim \Sigma(r_{\rm i}) (r/r_{\rm i})^{-\zeta}$, one can arrive at an analytical estimate of $\Omega_{\rm prec}$ given by eq.(43) in \citet{Fragile07}. 

\begin{figure}
    \centering
    \includegraphics[scale = 0.53]{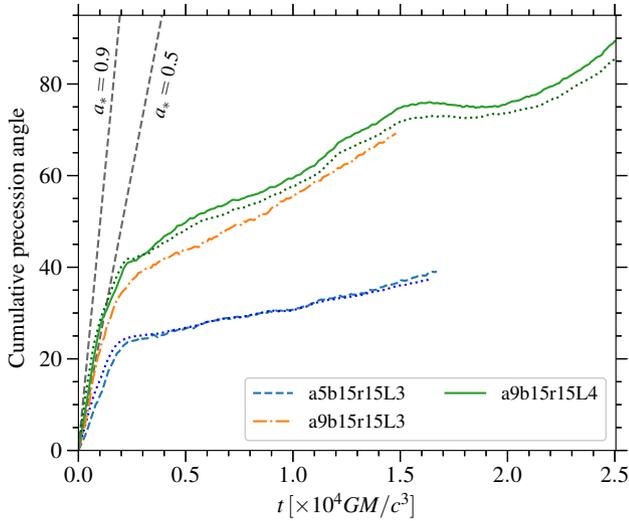}
    \caption{Time evolution of the cumulative precession angle over the range $5$-$15$\rg. The slopes of these curves give the precession rates. The grey, dashed lines represent the precession rates estimated from eq. (\ref{eq:prec}). The coloured, dotted curves represent the precession estimated using eq.~(\ref{eq:ang_evol}) for the simulations a9b15r15L4 and a5b15r15L3.}
    \label{fig:twist_comp}
\end{figure}

Fig.~\ref{fig:twist_comp} shows the cumulative precession angle for matter between 5 and 15\rg\, in our three simulations. We exclude the region inside $r=5$\rg, because tilted discs have effective inner radii well outside of the marginally stable limit \citep{Fragile09}. The expected precession rates [eq. (\ref{eq:prec})] for tori extending over this range of radii are represented by the grey dashed lines, with $\zeta=-6.1$ for the $a_*=0.9$ simulation and $-5.2$ for the $a_*=0.5$ one. We see that the predicted precession rates agree reasonably well with the early ($t \lesssim 1000 r_{\rm g}/c$) behavior of the simulations. However, after $t\approx 2000 r_{\rm g}/c$, which is roughly when the bending wave reaches the transition region ($15\lesssim r \lesssim 20$\rg, see Fig.~\ref{fig:twist}), the slopes of the cumulative precession curves change dramatically. The precession rate decreases by more than 95 percent to a rate comparable to what eq. (\ref{eq:prec}) would give if $r_{\rm o}$ were $40 r_{\rm g}$. This establishes the key results of our paper: \textit{the outer thin disc clearly affects the precession of the inner thick disc by reducing its rate}. It is encouraging that we find nearly identical precession rates in the low- and high-resolution, $a_*=0.9$ simulations, suggesting that this conclusion is robust. To ensure that the decrease in the precession rate is due to the outer thin disc, we performed another simulation identical to a9b15r15L4, except that the outer thin disc is absent, and found that it precesses continuously at the rate predicted by eq.~\ref{eq:prec}.

The decrease in the precession rate in the truncated disc simulations can be attributed to the exchange of angular momentum between the thick and thin discs. We discuss how to estimate this effect in Appendix \ref{appA}. In particular, if we include both the torque and flux terms in eq.~(\ref{eq:ang_evol}), then we find remarkable agreement between eq. (\ref{eq:dgdt}) and the simulation results, as shown by the dotted curves in Fig.~\ref{fig:twist_comp}. It is worth pointing out that the if we exclude the flux terms in eq.~(\ref{eq:ang_evol}), we recover the expected precession rates from eq. (\ref{eq:prec}).

\section{Discussion}
\label{discussion}
Over the last decade, the Lense-Thirring precession model by \citet{Ingram09} has gained wide popularity for explaining the type-C QPOs observed in X-ray binary systems. The dependence of the precession frequency on the outer radius of the torus in this model, facilitates an independent measurement of the truncation radius when compared to observations. According to  \citet{Ingram09}, the  estimated frequencies for an isolated torus with $r_{\rm i}$ between $5-10$\rg~ and $r_{\rm o}$ varying from $10-50$\rg~ can produce frequencies in the observed range of $0.01-10$~Hz. However, as we showed in the previous section, the presence of a thin disc outside the hot, thick flow decreases the precession frequency by about 95 percent. This has direct implications for the truncation radius estimation.

For a typical black hole mass of $10\,M_{\odot}$, the precession frequencies resulting from our simulations are roughly 0.14~Hz and 0.05~Hz for $a_* = 0.9$ and 0.5, respectively. These frequencies fall near the lower end of the type-C QPO range \citep{Remillard06}. To reach the upper end of the frequency range requires decreasing the size of the precessing torus. In this way, the slower precession frequencies noted in this work actually help relieve some of the tensions between the Lense-Thirring precession model and some type-C QPO observations. One such example is the 2.2~Hz QPO observed in the black hole binary system GRS~1915+105 \citep{Nathan22}. According to eq. (\ref{eq:prec}), the outer edge of an isolated torus precessing around a $10.1\,M_{\odot}$ black hole with $a_* = 0.998$ would need to be around $41$\rg~ to produce a frequency this low. However, the inferred truncation radius from phase-resolved spectroscopy of this particular QPO is $1.43^{+0.01}_{-0.02}$\rg~\citep{Nathan22}. One way this conflict could be resolved is if the outer, thin disc decreases the Lense-Thirring precession frequency of the inner torus as found in our work. Assuming that the percentage decrease in the precession frequency does not vary with the truncation radius, we can match the 2.2~Hz frequency with an outer edge now only at $\approx 11$\rg. This is still in conflict with the phase-resolved spectroscopy results, but moves things considerably in the right direction. Of course, we do not know yet how the precession rate will vary as the truncation radius changes, something that will need to be explored in future studies.

We conclude that in a truncated-disc geometry, the rate of Lense-Thirring precession of the hot, thick flow around a black hole can be significantly reduced from the isolated case by the advected angular momentum fluxes from the surrounding thin disc. In principle, this result could potentially be relevant to other astrophysical phenomena based on Lense-Thirring precession, such as tilted discs in tidal disruption events (TDEs), where the exchange of angular momentum between the incoming debris stream from the disrupted star and the precessing disc around the supermassive black hole could potentially slow down the precession or even halt it if the fluxes are strong enough; this is something that will be studied with future simulations. 

\section*{Acknowledgements}
We want to thank Pavel Ivanov for helpful discussions regarding this work. This research was partly supported by the Polish NCN grants 2018/29/N/ST9/02139, 2019/33/B/ST9/01564 and Polish National Agency for Academic Exchange grant PPN/IWA/2018/1/00099/U/0001. Simulations in this work were performed using the Cyfronet Prometheus cluster, part of the PL-Grid computing network and the Max Planck Computing \& Data Facility. This work also used the Extreme Science and Engineering Discovery Environment (XSEDE), which is supported by National Science Foundation grant (NSF) number ACI-1548562. PCF gratefully acknowledges the support of the NSF through grants AST1616185, PHY-1748958, and AST-1907850. This work was performed in part at the Aspen Center for Physics, which is supported by NSF grant PHY-1607611.

%%%%%%%%%%%%%%%%%%%%%%%%%%%%%%%%%%%%%%%%%%%%%%%%%%
\section*{Data Availability}
The data underlying this paper will be shared upon reasonable request to the corresponding author. 

\bibliographystyle{mnras}
\bibliography{ThinThick} %

\begin{thebibliography}{}
\makeatletter
\relax
\def\mn@urlcharsother{\let\do\@makeother \do\$\do\&\do\#\do\^\do\_\do\%\do\~}
\def\mn@doi{\begingroup\mn@urlcharsother \@ifnextchar [ {\mn@doi@}
  {\mn@doi@[]}}
\def\mn@doi@[#1]#2{\def\@tempa{#1}\ifx\@tempa\@empty \href
  {http://dx.doi.org/#2} {doi:#2}\else \href {http://dx.doi.org/#2} {#1}\fi
  \endgroup}
\def\mn@eprint#1#2{\mn@eprint@#1:#2::\@nil}
\def\mn@eprint@arXiv#1{\href {http://arxiv.org/abs/#1} {{\tt arXiv:#1}}}
\def\mn@eprint@dblp#1{\href {http://dblp.uni-trier.de/rec/bibtex/#1.xml}
  {dblp:#1}}
\def\mn@eprint@#1:#2:#3:#4\@nil{\def\@tempa {#1}\def\@tempb {#2}\def\@tempc
  {#3}\ifx \@tempc \@empty \let \@tempc \@tempb \let \@tempb \@tempa \fi \ifx
  \@tempb \@empty \def\@tempb {arXiv}\fi \@ifundefined
  {mn@eprint@\@tempb}{\@tempb:\@tempc}{\expandafter \expandafter \csname
  mn@eprint@\@tempb\endcsname \expandafter{\@tempc}}}

\bibitem[\protect\citeauthoryear{{Anninos}, {Fragile}  \&
  {Salmonson}}{{Anninos} et~al.}{2005}]{Anninos05}
{Anninos} P.,  {Fragile} P.~C.,   {Salmonson} J.~D.,  2005, \mn@doi [\apj]
  {10.1086/497294}, \href
  {https://ui.adsabs.harvard.edu/abs/2005ApJ...635..723A} {635, 723}

\bibitem[\protect\citeauthoryear{{Begelman} \& {Armitage}}{{Begelman} \&
  {Armitage}}{2014}]{BA2014}
{Begelman} M.~C.,  {Armitage} P.~J.,  2014, \mn@doi [\apjl]
  {10.1088/2041-8205/782/2/L18}, \href
  {https://ui.adsabs.harvard.edu/abs/2014ApJ...782L..18B} {782, L18}

\bibitem[\protect\citeauthoryear{{Belloni}}{{Belloni}}{2010}]{B2010}
{Belloni} T.~M.,  2010, in {Belloni} T.,  ed., , Vol.~794, Lecture Notes in
  Physics, Berlin Springer Verlag.
p.~53

\bibitem[\protect\citeauthoryear{{Beloborodov}}{{Beloborodov}}{1999}]{B1999}
{Beloborodov} A.~M.,  1999, \mn@doi [\apjl] {10.1086/311810}, \href
  {https://ui.adsabs.harvard.edu/abs/1999ApJ...510L.123B} {510, L123}

\bibitem[\protect\citeauthoryear{{Chakrabarti}}{{Chakrabarti}}{1985}]{C1985}
{Chakrabarti} S.~K.,  1985, \mn@doi [\apj] {10.1086/162755}, \href
  {https://ui.adsabs.harvard.edu/abs/1985ApJ...288....1C} {288, 1}

\bibitem[\protect\citeauthoryear{{Chakrabarti} \& {Titarchuk}}{{Chakrabarti} \&
  {Titarchuk}}{1995}]{CT95}
{Chakrabarti} S.,  {Titarchuk} L.~G.,  1995, \mn@doi [\apj] {10.1086/176610},
  \href {https://ui.adsabs.harvard.edu/abs/1995ApJ...455..623C} {455, 623}

\bibitem[\protect\citeauthoryear{{Colella} \& {Woodward}}{{Colella} \&
  {Woodward}}{1984}]{Colella1984}
{Colella} P.,  {Woodward} P.~R.,  1984, \mn@doi [Journal of Computational
  Physics] {10.1016/0021-9991(84)90143-8}, \href
  {https://ui.adsabs.harvard.edu/abs/1984JCoPh..54..174C} {54, 174}

\bibitem[\protect\citeauthoryear{{Eardley}, {Lightman}  \& {Shapiro}}{{Eardley}
  et~al.}{1975}]{EL1975}
{Eardley} D.~M.,  {Lightman} A.~P.,   {Shapiro} S.~L.,  1975, \mn@doi [\apjl]
  {10.1086/181871}, \href
  {https://ui.adsabs.harvard.edu/abs/1975ApJ...199L.153E} {199, L153}

\bibitem[\protect\citeauthoryear{{Esin}, {McClintock}  \& {Narayan}}{{Esin}
  et~al.}{1997}]{EMN1997}
{Esin} A.~A.,  {McClintock} J.~E.,   {Narayan} R.,  1997, \mn@doi [\apj]
  {10.1086/304829}, \href
  {https://ui.adsabs.harvard.edu/abs/1997ApJ...489..865E} {489, 865}

\bibitem[\protect\citeauthoryear{{Fabian} et~al.,}{{Fabian}
  et~al.}{2009}]{Fabian2009}
{Fabian} A.~C.,  et~al., 2009, \mn@doi [\nat] {10.1038/nature08007}, \href
  {https://ui.adsabs.harvard.edu/abs/2009Natur.459..540F} {459, 540}

\bibitem[\protect\citeauthoryear{{Fender}, {Belloni}  \& {Gallo}}{{Fender}
  et~al.}{2004}]{FBG2004}
{Fender} R.~P.,  {Belloni} T.~M.,   {Gallo} E.,  2004, \mn@doi [\mnras]
  {10.1111/j.1365-2966.2004.08384.x}, \href
  {https://ui.adsabs.harvard.edu/abs/2004MNRAS.355.1105F} {355, 1105}

\bibitem[\protect\citeauthoryear{{Fragile}}{{Fragile}}{2009}]{Fragile09}
{Fragile} P.~C.,  2009, \mn@doi [\apjl] {10.1088/0004-637X/706/2/L246}, \href
  {https://ui.adsabs.harvard.edu/abs/2009ApJ...706L.246F} {706, L246}

\bibitem[\protect\citeauthoryear{{Fragile}, {Blaes}, {Anninos}  \&
  {Salmonson}}{{Fragile} et~al.}{2007}]{Fragile07}
{Fragile} P.~C.,  {Blaes} O.~M.,  {Anninos} P.,   {Salmonson} J.~D.,  2007,
  \mn@doi [\apj] {10.1086/521092}, \href
  {https://ui.adsabs.harvard.edu/abs/2007ApJ...668..417F} {668, 417}

\bibitem[\protect\citeauthoryear{{Fragile}, {Wilson}  \& {Rodriguez}}{{Fragile}
  et~al.}{2012a}]{FW12}
{Fragile} P.~C.,  {Wilson} J.,   {Rodriguez} M.,  2012a, \mn@doi [\mnras]
  {10.1111/j.1365-2966.2012.21222.x}, \href
  {https://ui.adsabs.harvard.edu/abs/2012MNRAS.424..524F} {424, 524}

\bibitem[\protect\citeauthoryear{{Fragile}, {Wilson}  \& {Rodriguez}}{{Fragile}
  et~al.}{2012b}]{Fragile12}
{Fragile} P.~C.,  {Wilson} J.,   {Rodriguez} M.,  2012b, \mn@doi [\mnras]
  {10.1111/j.1365-2966.2012.21222.x}, \href
  {https://ui.adsabs.harvard.edu/abs/2012MNRAS.424..524F} {424, 524}

\bibitem[\protect\citeauthoryear{{Galeev}, {Rosner}  \& {Vaiana}}{{Galeev}
  et~al.}{1979}]{GRV1979}
{Galeev} A.~A.,  {Rosner} R.,   {Vaiana} G.~S.,  1979, \mn@doi [\apj]
  {10.1086/156957}, \href
  {https://ui.adsabs.harvard.edu/abs/1979ApJ...229..318G} {229, 318}

\bibitem[\protect\citeauthoryear{{Haardt} \& {Maraschi}}{{Haardt} \&
  {Maraschi}}{1993}]{HFM1993}
{Haardt} F.,  {Maraschi} L.,  1993, \mn@doi [\apj] {10.1086/173020}, \href
  {https://ui.adsabs.harvard.edu/abs/1993ApJ...413..507H} {413, 507}

\bibitem[\protect\citeauthoryear{Harten, Lax  \& Leer}{Harten
  et~al.}{1983}]{Harten1983}
Harten A.,  Lax P.~D.,   Leer B.~V.,  1983, SIAM Review, 25, 35

\bibitem[\protect\citeauthoryear{{Ingram} \& {Done}}{{Ingram} \&
  {Done}}{2011}]{ID2011}
{Ingram} A.,  {Done} C.,  2011, \mn@doi [\mnras]
  {10.1111/j.1365-2966.2011.18860.x}, \href
  {https://ui.adsabs.harvard.edu/abs/2011MNRAS.415.2323I} {415, 2323}

\bibitem[\protect\citeauthoryear{{Ingram} \& {Motta}}{{Ingram} \&
  {Motta}}{2019}]{IM2019}
{Ingram} A.~R.,  {Motta} S.~E.,  2019, \mn@doi [\nar]
  {10.1016/j.newar.2020.101524}, \href
  {https://ui.adsabs.harvard.edu/abs/2019NewAR..8501524I} {85, 101524}

\bibitem[\protect\citeauthoryear{{Ingram}, {Done}  \& {Fragile}}{{Ingram}
  et~al.}{2009}]{Ingram09}
{Ingram} A.,  {Done} C.,   {Fragile} P.~C.,  2009, \mn@doi [\mnras]
  {10.1111/j.1745-3933.2009.00693.x}, \href
  {https://ui.adsabs.harvard.edu/abs/2009MNRAS.397L.101I} {397, L101}

\bibitem[\protect\citeauthoryear{{Ingram}, {van der Klis}, {Middleton}, {Done},
  {Altamirano}, {Heil}, {Uttley}  \& {Axelsson}}{{Ingram}
  et~al.}{2016}]{Ingram2016}
{Ingram} A.,  {van der Klis} M.,  {Middleton} M.,  {Done} C.,  {Altamirano} D.,
   {Heil} L.,  {Uttley} P.,   {Axelsson} M.,  2016, \mn@doi [\mnras]
  {10.1093/mnras/stw1245}, \href
  {https://ui.adsabs.harvard.edu/abs/2016MNRAS.461.1967I} {461, 1967}

\bibitem[\protect\citeauthoryear{{Kato} \& {Fukue}}{{Kato} \&
  {Fukue}}{1980}]{KF80}
{Kato} S.,  {Fukue} J.,  1980, \pasj, \href
  {https://ui.adsabs.harvard.edu/abs/1980PASJ...32..377K} {32, 377}

\bibitem[\protect\citeauthoryear{{Liska}, {Hesp}, {Tchekhovskoy}, {Ingram},
  {van der Klis}  \& {Markoff}}{{Liska} et~al.}{2018}]{Liska18}
{Liska} M.,  {Hesp} C.,  {Tchekhovskoy} A.,  {Ingram} A.,  {van der Klis} M.,
  {Markoff} S.,  2018, \mn@doi [\mnras] {10.1093/mnrasl/slx174}, \href
  {https://ui.adsabs.harvard.edu/abs/2018MNRAS.474L..81L} {474, L81}

\bibitem[\protect\citeauthoryear{{Liu}, {Taam}, {Meyer-Hofmeister}  \&
  {Meyer}}{{Liu} et~al.}{2007}]{LTMM2007}
{Liu} B.~F.,  {Taam} R.~E.,  {Meyer-Hofmeister} E.,   {Meyer} F.,  2007,
  \mn@doi [\apj] {10.1086/522619}, \href
  {https://ui.adsabs.harvard.edu/abs/2007ApJ...671..695L} {671, 695}

\bibitem[\protect\citeauthoryear{{Martocchia} \& {Matt}}{{Martocchia} \&
  {Matt}}{1996}]{MM1996}
{Martocchia} A.,  {Matt} G.,  1996, \mn@doi [\mnras] {10.1093/mnras/282.4.L53},
  \href {https://ui.adsabs.harvard.edu/abs/1996MNRAS.282L..53M} {282, L53}

\bibitem[\protect\citeauthoryear{{McClintock} \& {Remillard}}{{McClintock} \&
  {Remillard}}{2006}]{MR2006}
{McClintock} J.~E.,  {Remillard} R.~A.,  2006, in , Vol.~39, Compact stellar
  X-ray sources.
pp 157--213

\bibitem[\protect\citeauthoryear{{McKinney}}{{McKinney}}{2006}]{McKinney06}
{McKinney} J.~C.,  2006, \mn@doi [\mnras] {10.1111/j.1365-2966.2006.10256.x},
  \href {https://ui.adsabs.harvard.edu/abs/2006MNRAS.368.1561M} {368, 1561}

\bibitem[\protect\citeauthoryear{{Motta}}{{Motta}}{2016}]{Motta16}
{Motta} S.~E.,  2016, \mn@doi [Astronomische Nachrichten]
  {10.1002/asna.201612320}, \href
  {https://ui.adsabs.harvard.edu/abs/2016AN....337..398M} {337, 398}

\bibitem[\protect\citeauthoryear{{Motta}, {Homan}, {Mu{\~n}oz Darias},
  {Casella}, {Belloni}, {Hiemstra}  \& {M{\'e}ndez}}{{Motta}
  et~al.}{2012}]{Motta12}
{Motta} S.,  {Homan} J.,  {Mu{\~n}oz Darias} T.,  {Casella} P.,  {Belloni}
  T.~M.,  {Hiemstra} B.,   {M{\'e}ndez} M.,  2012, \mn@doi [\mnras]
  {10.1111/j.1365-2966.2012.22037.x}, \href
  {https://ui.adsabs.harvard.edu/abs/2012MNRAS.427..595M} {427, 595}

\bibitem[\protect\citeauthoryear{{Motta}, {Casella}, {Henze},
  {Mu{\~n}oz-Darias}, {Sanna}, {Fender}  \& {Belloni}}{{Motta}
  et~al.}{2015}]{Motta2015}
{Motta} S.~E.,  {Casella} P.,  {Henze} M.,  {Mu{\~n}oz-Darias} T.,  {Sanna} A.,
   {Fender} R.,   {Belloni} T.,  2015, \mn@doi [\mnras]
  {10.1093/mnras/stu2579}, \href
  {https://ui.adsabs.harvard.edu/abs/2015MNRAS.447.2059M} {447, 2059}

\bibitem[\protect\citeauthoryear{{Nathan} et~al.,}{{Nathan}
  et~al.}{2022}]{Nathan22}
{Nathan} E.,  et~al., 2022, \mn@doi [\mnras] {10.1093/mnras/stab3803}, \href
  {https://ui.adsabs.harvard.edu/abs/2022MNRAS.511..255N} {511, 255}

\bibitem[\protect\citeauthoryear{{Parker} et~al.,}{{Parker}
  et~al.}{2015}]{Parker2015}
{Parker} M.~L.,  et~al., 2015, \mn@doi [\apj] {10.1088/0004-637X/808/1/9},
  \href {https://ui.adsabs.harvard.edu/abs/2015ApJ...808....9P} {808, 9}

\bibitem[\protect\citeauthoryear{{Pringle}}{{Pringle}}{1992}]{Pringle92}
{Pringle} J.~E.,  1992, \mn@doi [\mnras] {10.1093/mnras/258.4.811}, \href
  {https://ui.adsabs.harvard.edu/abs/1992MNRAS.258..811P} {258, 811}

\bibitem[\protect\citeauthoryear{{Remillard} \& {McClintock}}{{Remillard} \&
  {McClintock}}{2006}]{Remillard06}
{Remillard} R.~A.,  {McClintock} J.~E.,  2006, \mn@doi [\araa]
  {10.1146/annurev.astro.44.051905.092532}, \href
  {https://ui.adsabs.harvard.edu/abs/2006ARA&A..44...49R} {44, 49}

\bibitem[\protect\citeauthoryear{{Rodriguez}, {Varni{\`e}re}, {Tagger}  \&
  {Durouchoux}}{{Rodriguez} et~al.}{2002}]{RVT2002}
{Rodriguez} J.,  {Varni{\`e}re} P.,  {Tagger} M.,   {Durouchoux} P.,  2002,
  \mn@doi [\aap] {10.1051/0004-6361:20000524}, \href
  {https://ui.adsabs.harvard.edu/abs/2002A&A...387..487R} {387, 487}

\bibitem[\protect\citeauthoryear{{Silbergleit}, {Wagoner}  \&
  {Ortega-Rodr{\'\i}guez}}{{Silbergleit} et~al.}{2001}]{SWM01}
{Silbergleit} A.~S.,  {Wagoner} R.~V.,   {Ortega-Rodr{\'\i}guez} M.,  2001,
  \mn@doi [\apj] {10.1086/318659}, \href
  {https://ui.adsabs.harvard.edu/abs/2001ApJ...548..335S} {548, 335}

\bibitem[\protect\citeauthoryear{{Spiteri} \& {Ruuth}}{{Spiteri} \&
  {Ruuth}}{2002}]{Spiteri02}
{Spiteri} R.~J.,  {Ruuth} S.~J.,  2002, SIAM Journal on Numerical Analysis, 40,
  469

\bibitem[\protect\citeauthoryear{{Stella} \& {Vietri}}{{Stella} \&
  {Vietri}}{1998}]{SV98}
{Stella} L.,  {Vietri} M.,  1998, \mn@doi [\apjl] {10.1086/311075}, \href
  {https://ui.adsabs.harvard.edu/abs/1998ApJ...492L..59S} {492, L59}

\bibitem[\protect\citeauthoryear{{Stella}, {Vietri}  \& {Morsink}}{{Stella}
  et~al.}{1999}]{SVM99}
{Stella} L.,  {Vietri} M.,   {Morsink} S.~M.,  1999, \mn@doi [\apjl]
  {10.1086/312291}, \href
  {https://ui.adsabs.harvard.edu/abs/1999ApJ...524L..63S} {524, L63}

\bibitem[\protect\citeauthoryear{{Svensson}}{{Svensson}}{1996}]{S1996}
{Svensson} R.,  1996, \aaps, \href
  {https://ui.adsabs.harvard.edu/abs/1996A&AS..120C.475S} {120, 475}

\bibitem[\protect\citeauthoryear{{Tagger} \& {Pellat}}{{Tagger} \&
  {Pellat}}{1999}]{TP99}
{Tagger} M.,  {Pellat} R.,  1999, \aap, \href
  {https://ui.adsabs.harvard.edu/abs/1999A&A...349.1003T} {349, 1003}

\bibitem[\protect\citeauthoryear{{Wagoner}}{{Wagoner}}{1999}]{W1999}
{Wagoner} R.~V.,  1999, \mn@doi [\physrep] {10.1016/S0370-1573(98)00104-5},
  \href {https://ui.adsabs.harvard.edu/abs/1999PhR...311..259W} {311, 259}

\bibitem[\protect\citeauthoryear{{White}, {Quataert}  \& {Blaes}}{{White}
  et~al.}{2019}]{WQB19}
{White} C.~J.,  {Quataert} E.,   {Blaes} O.,  2019, \mn@doi [\apj]
  {10.3847/1538-4357/ab089e}, \href
  {https://ui.adsabs.harvard.edu/abs/2019ApJ...878...51W} {878, 51}

\bibitem[\protect\citeauthoryear{{van den Eijnden}, {Ingram}, {Uttley},
  {Motta}, {Belloni}  \& {Gardenier}}{{van den Eijnden}
  et~al.}{2017}]{eijnden2017}
{van den Eijnden} J.,  {Ingram} A.,  {Uttley} P.,  {Motta} S.~E.,  {Belloni}
  T.~M.,   {Gardenier} D.~W.,  2017, \mn@doi [\mnras] {10.1093/mnras/stw2634},
  \href {https://ui.adsabs.harvard.edu/abs/2017MNRAS.464.2643V} {464, 2643}

\makeatother
\end{thebibliography}
%%%%%%%%%%%%%%%%%%%%%%%%%%%%%%%%%%%%%%%%%%%%%%%%%%

%%%%%%%%%%%%%%%%% APPENDICES %%%%%%%%%%%%%%%%%%%%%

\appendix

\section{Calculation of precession rate}
\label{appA}
Let $J_1$ and $J_2$ be orthogonal vector components of the angular momentum of the precessing torus projected onto the equatorial plane of the black hole. We can then define the precession angle, $\gamma$, of the torus extending from $r_{\rm i}$ to $r_{\rm o}$ as $\tan \gamma = J_2/J_1$. Differentiating with respect to time gives us the precession rate, 
\begin{eqnarray}
\frac{\partial \gamma}{\partial t} = \frac{1}{J_1^2+J_2^2}\left(J_1\frac{\partial J_2}{\partial t} -J_2\frac{\partial J_1}{\partial t} \right) ~.
\label{eq:dgdt}
\end{eqnarray}
All we need to do is compute the terms $J_i$ and $\partial J_i/\partial t$ (for $i=1,2$) from the simulation data. 

For $J_i$, we take
\begin{eqnarray}
J_i = \int_{r_{\rm i}}^{r_{\rm o}} {\rm d}r {\rm d}\theta {\rm d}\phi \sqrt{-g} l_i ~,
\end{eqnarray}
where $l_1 = l^x\cos\beta_0+l^z\sin \beta_0$ and $l_2 = l^y$; the vector components of the angular momentum density ($\bm{l}$) in Cartesian coordinates are $l^x = yT^{tz} - zT^{ty}$, $l^y = zT^{tx} - xT^{tz}$, and $l^z = xT^{ty} - yT^{tx}$; and $T^{\mu\nu} = \left( \rho h+2P_{\rm m}\right)u^{\mu}u^{\nu} + \left(P_{\rm g}+P_{\rm m}\right)g^{\mu \nu}-b^{\mu}b^{\nu}$ is the stress-energy tensor of the fluid in the MHD limit, where $h = 1+\epsilon+P_{\rm g}/\rho$ is the specific enthalpy, $\epsilon$ is the specific internal energy density, $P_{\rm m}$ is the magnetic pressure, $u^{\mu}$ is the fluid four-velocity, and $b^{\mu}$ is the four-vector  magnetic field measured by an observer comoving with the fluid.

To predict the time-derivatives of the angular momentum in a simple model, we use the angular momentum conservation equation \citep{Pringle92}
\begin{eqnarray}
    \frac{\partial \bm{L} }{\partial t} + \frac{1}{r}\frac{\partial}{\partial r}\left(r V^r \bm{L}\right) = \bm{\Omega}_{\rm LT}\times\bm{L} ~,
\end{eqnarray}
where $\bm{L}$ is the angular momentum vector per unit surface area and $V^r$ is the radial velocity. Technically, this equation was derived in the diffusive regime, but as we show, it works well even for our truncated discs. We have ignored the viscous terms in the above conservation equation, as we find them to be negligible when compared to the Lense-Thirring torque and angular momentum fluxes in our simulations. Integrating the vector components of the above equation over the surface area, we get

\begin{eqnarray}
\frac{\partial J_i}{\partial t} = \mp \int_{r_{\rm i}}^{r_{\rm o}} {\rm d}r {\rm d}\theta {\rm d}\phi \sqrt{-g} \Omega_{\rm LT} l_j -\left[ F_i\vert_{r_{\rm o}} -F_i\vert_{r_{\rm i}}\right],
\label{eq:ang_evol}
\end{eqnarray}
where the minus sign and $j=2$ are taken for $i=1$, and the plus sign and $j=1$ are taken for $i=2$. The first term on the right-hand-side is the usual Lense-Thirring torque term, and the terms within braces represent the angular momentum flux through the boundaries, $r_{\rm i}$ and $r_{\rm o}$, given by
\begin{eqnarray}
F_i =\int {\rm d}\theta {\rm d}\phi \sqrt{-g} l_i V^r ~. 
\end{eqnarray}

%%%%%%%%%%%%%%%%%%%%%%%%%%%%%%%%%%%%%%%%%%%%%%%%%%

% Don't change these lines
\bsp	% typesetting comment
\label{lastpage}
\end{document}